\documentclass{acmtog}

\usepackage{epsfig,endnotes}
\usepackage{color}
\usepackage{graphicx}
\usepackage{subfig}
\DeclareCaptionType{copyrightbox}
\usepackage{url}                % for URL's in biblio

\newcommand{\myurl}[1]{\url{#1}}                % Use url package or something else

\definecolor{gray}{RGB}{128,128,128}
\begin{document}

\title{Secure Deletion on Log-structured File Systems}
% FOR ANONYMOUS SUBMISSION
\author{
Joel Reardon, Claudio Marforio, Srdjan Capkun, David Basin \\
Department of Computer Science, ETH Zurich}
%\category{hm}{a}{a}
%\terms{Security}
%%\keywords{Privacy, Flash-memory} 
\maketitle

%\begin{bottomstuff}
%mysterious bottom stuff
%\end{bottomstuff}

\begin{abstract}
We address the problem of secure data deletion on log-structured file systems.
We focus on the YAFFS file system, widely used on  Android smartphones. 
We show that these systems provide no temporal guarantees on data deletion and 
that deleted data still persists for nearly  
44 hours with average phone use and 
indefinitely if the phone is not used after the deletion.
Furthermore, we show that file overwriting and encryption, methods commonly used for
secure deletion on block-structured file systems, 
do not ensure data deletion in log-structured file systems.

We propose three mechanisms for secure deletion on log-structured file
systems. \emph{Purging} is a user-level mechanism that guarantees secure deletion at the cost
of negligible device wear. \emph{Ballooning} is a user-level mechanism that
runs continuously and gives probabilistic improvements to secure deletion.  
\emph{Zero overwriting} is  a kernel-level mechanism that guarantees immediate secure
deletion without device wear. 
We implement these mechanisms on Nexus One smartphones and show that they
succeed in secure deletion and 
neither prohibitively 
reduce the longevity of the flash memory nor noticeably reduce the device's
battery lifetime. These techniques provide mobile phone 
users more confidence that data
they delete from their phones are indeed deleted.

%We also describe conditions under which our mechanisms can be applied
%to other flash file systems such as JFFS and propose a modification to UBI---a
%higher-level flash interface---that would implement Ballooning on any flash file
%system that uses it.

\end{abstract}

\section{Introduction}
Deleting a file from a storage medium serves two  purposes: it
reclaims storage and ensures that any sensitive information
contained in the file is no longer accessible. When done for the
latter purpose, it is critical that the file is \emph{securely} deleted,
meaning that its content does not persist on the storage medium after
deletion.

Secure deletion is almost always ignored in
file system design~\cite{ext2,yaffs,ubifs,ext4,logstructure,jffs}, largely for
performance reasons.  Typically, deletion is implemented as a rapid operation where a file is \emph{unlinked}, meaning its metadata
states that it is no longer present while the file's contents remain in
storage until overwritten by new data~\cite{survey}.
While many users expect that deleting messages will delete them, clearing the browser's history will clear it, and changing their
location will overwrite their previous location, in reality this information
remains on their devices without any guarantees of deletion.
Surveys of repurposed hard drives found that many contained private financial or
medical data that could be recovered with trivial forensic cost and
effort~\cite{survey}. 
However, secure deletion is not only important when media is repurposed; it
also enables users to protect the confidentiality of their data 
if their devices are compromised, stolen, or confiscated under a subpoena. In
the case of a subpoena, the user may 
be forced to disclose all passwords, keys, or other credentials that enable access to the data stored on the device; 
in such a scenario, users can only sanitize their device 
 before it is seized. 

Secure deletion is particularly important on modern smartphones,
as they increasingly 
store personal data such as the owner's private conversations, browsing
history, and location history.
%\footnote{California's Supreme Court ruled that 
%police may search the
%mobile phone of anyone who is arrested---provided they are carrying their
%phone at the time of arrest~\cite{california}.} 
Mobile phones further store business data, which for company policy or legal reasons should be 
deleted after some time elapses or should not be available at some geographic
locations, e.g., should be deleted before a phone owner leaves a certain
jurisdiction.
%\footnote{Due to legal constraints, personal and financial data of
%bank customers typically should not be available out of the country of their
%bank account even when encrypted.}
Currently, the only secure deletion option
available on the Android mobile phone is the \emph{factory reset}: a
procedure that securely erases all user data on the phone, 
returning the phone to its initial state. This is clearly inappropriate for users who
wish to selectively delete data, such as some emails, but retain other
data, such as their address books or installed applications.

Secure deletion mechanisms have been proposed for some widely-used block-structured and
journalled file systems, like ext2 and
ext3~\cite{ext2secdel,securemyths, purgefs}. These mechanisms typically modify
the kernel and enforce that 
when a file is marked for deletion it is overwritten with arbitrary data.
Another mechanism that was proposed for preserving data confidentiality under
device seizure is full-drive
encryption, where the decryption key is generated from user's password upon
system boot. This mechanism is effective but has limitations: users can be
coerced or legally bound to disclose their credentials, or might be legally
obliged to delete data.

In this work, we address secure deletion on modern smartphones
with a focus on the file system YAFFS~\cite{yaffs}. 
Unlike ext2 and ext3, YAFFS is a log-structured file system developed
specifically for flash memory storage. Android phones' internal memory 
uses YAFFS to store data such as browsing cache, maps cache,
names of nearby wireless networks, GPS location data, SMS messages, electronic
mails, and telephone call listings. 

We analyze how deletion is
performed in YAFFS and show that log-structured file systems provide no
temporal guarantees on data deletion; deleted data persists
for around 44 hours with average phone use (Nexus
One~\cite{nexus}) and indefinitely if the phone is not used after the
deletion.  Furthermore, we show that mechanisms such as file
overwriting or encryption of individual files, proposed for data
deletion on block-structured file systems~\cite{shred,gpg} do
not ensure data deletion in log-structured file systems.
Namely, overwriting a file in log-structured file systems
simply writes a new version of a file, but does not remove  the
original copy. Similarly, when a file is encrypted, the ciphertext
will be written to a new location, but the plaintext will remain on
the flash drive until storage space is needed and garbage
collection is invoked. 

We propose three mechanisms for secure deletion
in YAFFS, two programs at user-level and one kernel-level file system change. 
The two user-level mechanisms are \emph{purging}, which
provides guaranteed rapid deletion of all data previously marked to be
deleted, and \emph{ballooning}, a continuous operation that reduces the
expected time that any piece of deleted data remains on the medium. The third mechanism is 
\emph{zero overwriting}, a kernel-level file system change that
securely deletes any data the moment it is removed from the file system.

We implement these mechanisms for the Nexus One smartphone and show that they
neither prohibitively 
reduce the longevity of the flash memory nor noticeably reduce the device's battery
lifetime. The purging operation occupies the phone
for half a minute, but it can be configured to
run during the phone's idle time. Ballooning provides a trade off between the time
until data is securely erased and the resulting wear on the flash memory. 
Zero overwriting securely deletes data immediately without imposing any
additional wear on the device. However, it has the drawback of requiring
kernel-level modifications and may not be suitable for all flash memories. 

Finally, we discuss the conditions under which our mechanisms can be applied to other log-structured 
file systems such as JFFS, and propose a modification to UBI---a 
higher-level flash interface---that would implement ballooning on any flash
file system that makes use of it.

The rest of this paper is organized as follows. In Section~2 we give background
on flash memory and file systems. In Section~3 we examine the current state
of secure deletion in YAFFS. In Sections~4 and~5 we present our three solutions,
along with experimental results.
In Section~6 we discuss related work and in Section~7 we discuss
generalizations of our approach and introduce future work.

\newpage

\section{System Model and Background}

\paragraph{System Model}
We consider a scenario in which users have
data on their mobile device that they wish to securely delete. This includes
cache files that should be \emph{continually} deleted, such as their location
histories as encoded by the names 
of the wireless networks and cellular base stations with which  their devices
communicated.

The adversary that we consider in this paper is the \emph{coercion attacker}, the
strongest attacker for the data deletion problem. The attacker can---at any
moment---both obtain the user's device and compel the user to reveal any secret
keys and passphrases~\cite{porter}.  The unpredictable nature of the attack 
prevents the user from performing any phone sanitization procedure
before disclosure. This differentiates the secure deletion problem from data deletion in the
context of repurposed hardware~\cite{survey}.

This strong attacker model also means that simple solutions to preserving data
confidentiality under device compromise will fail. Encryption of all the data
that is written onto the device would not work 
since the adversary will
be given all our encryption keys.
The use of factory reset would not be practical
as the unpredictable compromise time would require erasing
the entire phone's memory with such
frequency that little useful data could reside on the device.

We therefore need novel solutions to this problem. We consider solutions at two
levels of system integration: 
user-level and kernel-level. User-level means that our
application can only perform actions that a normal application installed from
the marketplace can perform. This mode of access is greatly limited: an
application's interaction with the file system consists solely of the creation
and deletion of its own local files. It cannot change the file system's behaviour 
in any
way to achieve secure deletion. Kernel-level access is much less
limited: it assumes that arbitrary changes can be made to the file system and a new
kernel can be installed on the device.

\paragraph{Log-structured File Systems}
A log-structured file system differs from a traditional block-based
file system (such as FAT or ext2) in that the entire file system is stored as a chronological record
of changes from the initial empty state.  As files are written, new fixed-size
\emph{chunks} are
appended to the log indicating the resulting change; a chunk can store either
a file's header or some  data, and is always added to the log's end. 
The file system maintains in RAM information on where the newest version of
each header and data chunk can be found.

Log-structured file systems complicate secure deletion because the
traditional approach of overwriting a file with new content simply
appends a second version of the file, while the first still remains in the
log's history. Similarly,
encrypting a file will also just append a new encrypted version of that
file, while the plaintext remains in the log.

Data is only removed from a log-structured file system
during \emph{garbage collection}. The garbage collector operates at the \emph{erase block}
level, which has a  larger granularity than a chunk.
The file system examines the wasted space in an erase block and
the total remaining free space when deciding whether to
garbage collect the erase block.

\paragraph{YAFFS}

Yet Another Flash File System (YAFFS) is a flash-based log-structured file
system that is notably used for the internal memory  of Android mobile phones.
YAFFS allocates memory by selecting an unused erase block and allocating
sequentially the numbered chunks in that block. When the block contains no more empty
chunks, a new block is selected for allocation by searching for an empty
block. YAFFS  searches for empty blocks sequentially, wrapping cyclically when necessary, 
by the erase block number as defined by the
physical layout of memory on the storage medium. It begins its search from the
last allocated block and returns the first empty block it finds. When allocating
a block  reduces the total number of empty blocks in the system below the
minimum threshold, then blocks containing wasted space are
compacted to reclaim storage. If there is no block that can be
compacted, that is, there is not a single unneeded chunk stored on the medium, then
YAFFS reports the file system as full and fails to allocate a block.

Garbage collection in YAFFS is either initiated by a thread that
performs system maintenance, or takes place during write operations. Usually, only a few
chunks are copied at a time, whereby the work to copy a block is amortized over many
write operations. If the file system contains too few free blocks then a
more aggressive garbage collection is performed. In this case, blocks with
less deleted space are collected, and the procedure  continues until
the entire block can be reclaimed.

\begin{figure}[t]
\centering
\includegraphics[scale=.8]{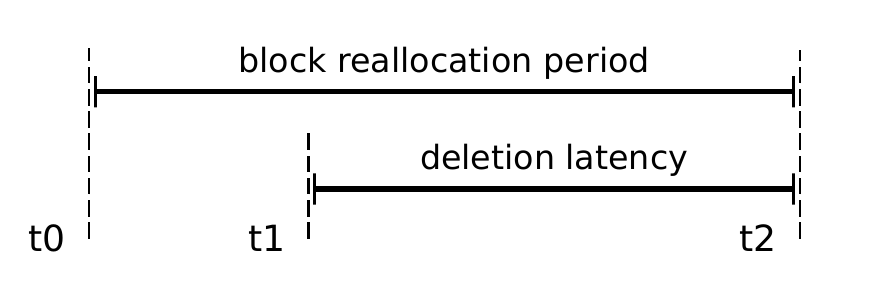}
\caption{\small A lifetime of stored data. 
\label{f:deletion_time}\normalsize}
\end{figure}

Figure~\ref{f:deletion_time} shows the lifetime for stored data. At time $t_0$
the block is
allocated and data is written onto it soon after. At time $t_1$ the data is
deleted. At time $t_2$ the block is reallocated, thus removing the data from
the medium. The difference $t_2-t_1$ is called the deletion latency, and $t_2-t_0$ is called
the block reallocation period.

\paragraph{Flash Memory}
Flash memory is a non-volatile storage medium consisting of an array of
electrical components that store information. 
The contents of flash memory cannot be
altered in place, but rather an erase procedure must be performed on a
larger granularity than reading or writing.
Flash erasure is costly: its increased
voltage requirement precipitate the wearing out of the medium.
Erase blocks can only handle a finite
number of erasure operations---roughly $10^4$ to $10^5$~\cite{flashlife}---before
becoming unusable. 

Flash file systems are typically log-structured for two
reasons. First, the large erase granularity of flash memory maps exactly to
the garbage collector's erase blocks in a log-structured file system. Second,
log-structured file systems do not require in-place updates for data; this
is well-suited to flash memory's inability to perform in-place updates.

%\paragraph{JFFS}

%The Journalling Flash File System is a pure log-structured file system. The log
%is stored contiguously in the physical layout of the blocks of flash memory.
%It has the interesting property of perfect wear-leveling; when the head
%approaches the tail, then the block pointed to by the tail is erased and the tail
%advances. This occurs even if no space was wasted on the tail block---all the
%valid data is copied to the head and the tail is reclaimed. Erasing a block of
%data thus requires waiting until the head of the log reaches that block.
%Mounting is
%also expensive, as it requires every entry in the log to be scanned and
%interpreted to build the appropriate data structures in RAM for data
%access.

%The second version of JFFS added  data compression to
%flash nodes and ameliorated the linear garbage collection. 
%In JFFS2, blocks are no longer allocated sequentially by their physical
%layout, but can be allocated arbitrarily.
%Sequence numbers are written when a block is allocated so they can be ordered
%chronologically during mounting.
%When allocating a new block for the head of the log, the algorithm selects the
%block with the most wasted space: preferably an empty block, and otherwise a
%nearly-empty block. (For wear-leveling, it infrequently selects a random block.)  

\section{Data Deletion Guarantees in \\Existing YAFFS Systems}

In this section, we investigate data persistence on Android
phones.
We examine the time that it takes for one erase block to be
reclaimed after being marked for deletion. In particular,
we measure the average
and worst-case data deletion latency for specific
devices, application configurations and usage patterns. 
To measure the average time taken
for block reallocation, which implies the deletion of any
data previously stored on the block, we instrument the file system at the
kernel level to log block allocation information. 

Our results show the existence of a large deletion latency, where data that a
user may believe to be deleted in reality remains accessible on the mobile
phone. This motivates our secure deletion solutions in the next sections.

\subsection{Instrumented YAFFS}
We built a modified version of the YAFFS Linux kernel module that logs
data about block allocations and chunk writes.  We log whenever a new
block is allocated, which signals that the block is now empty and
that whatever data was previously on the block has been erased (or
moved). We also log every write operation: both of file headers and
of file data. This allows us to determine how often writes occur, in
which chunks they occur, and when files are deleted.

During block allocations, we log the system time in microseconds, the
unique physical block number (in our case, ranging from 1 to
1570), the block's sequence number, and the number of free chunks
and erased blocks according to YAFFS's statistics. We also log the
file system's partition name to demultiplex the data, as the Android phone has multiple YAFFS
partitions.

Logging every chunk write gives us a fine-grained view of the system's writing
and deleting behaviour. We log the system time in microseconds,
the chunk's physical location, the operating system's owner of
the file, the block on which it is written, the type of data being
written (i.e., a file, a directory, a header, etc.), the file id, and where in the
file the data is being written.

With the collected information, we can determine how much data is
written to the file system, and the timing and frequency of block
erasures. We can also log the time when we write a particular file to
the file system, which we cross-reference in our logs to determine the
block number on which it resides. Given this information, along with
the time when each block is erased and the time the data was marked for
deletion by the user, we can compute the deletion latency (cf.~Figure~\ref{f:deletion_time}).

By logging the ownership of chunks, we can also determine the distinct
writing patterns of different running applications. We will later use
this to construct profiles that model different scenarios in our
simulated environment.

\subsection{Deletion Latency on Android}

\label{s:deletion_android}

To understand the severity of the existing problem with current
implementations, we examine in detail the deletion latency
on an Android phone. First, we focus on a subset of
applications that could be used daily on a smart phone to determine deletion
latency when using only such applications. We then continue to use the phone
throughout our daily routine to find out, on average, how long data
remains ``alive'' on the system before being erased. The data we collected on
the phone's writing patterns was later used to simulate an Android mobile phone.

The system under test is a Nexus One running the latest Android OS
(2.2.1) under what can be considered normal daily use: browsing the
web, saving images, listening to music, writing and receiving SMS
messages, and making calls. 
To understand the writing patterns of some commonly used applications,
we let a user use the phone's browser, maps and gallery
applications plus a popular game found on the Android Market. The user
used the phone unaware of the test, thereby eliminating
any bias which could be introduced by knowing which system properties
were examined.

\paragraph{Writing Patterns}
For the \emph{browser} test, the user surfed the web for approximately
8 minutes, performing activities such as logging
into a university website, getting weather forecasts, and searching
for images. For the \emph{maps} case, the user interacted with the
application for approximately 6 minutes, searching for a particular
destination, looking at its ``street view'' and calculating a route to
it. The \emph{game} and \emph{gallery} examples ran for
approximately 4-5 minutes each.

\begin{figure}[tb!]
  \centering
  \subfloat[][\emph{Browser} application traces]{\includegraphics[scale=.35]{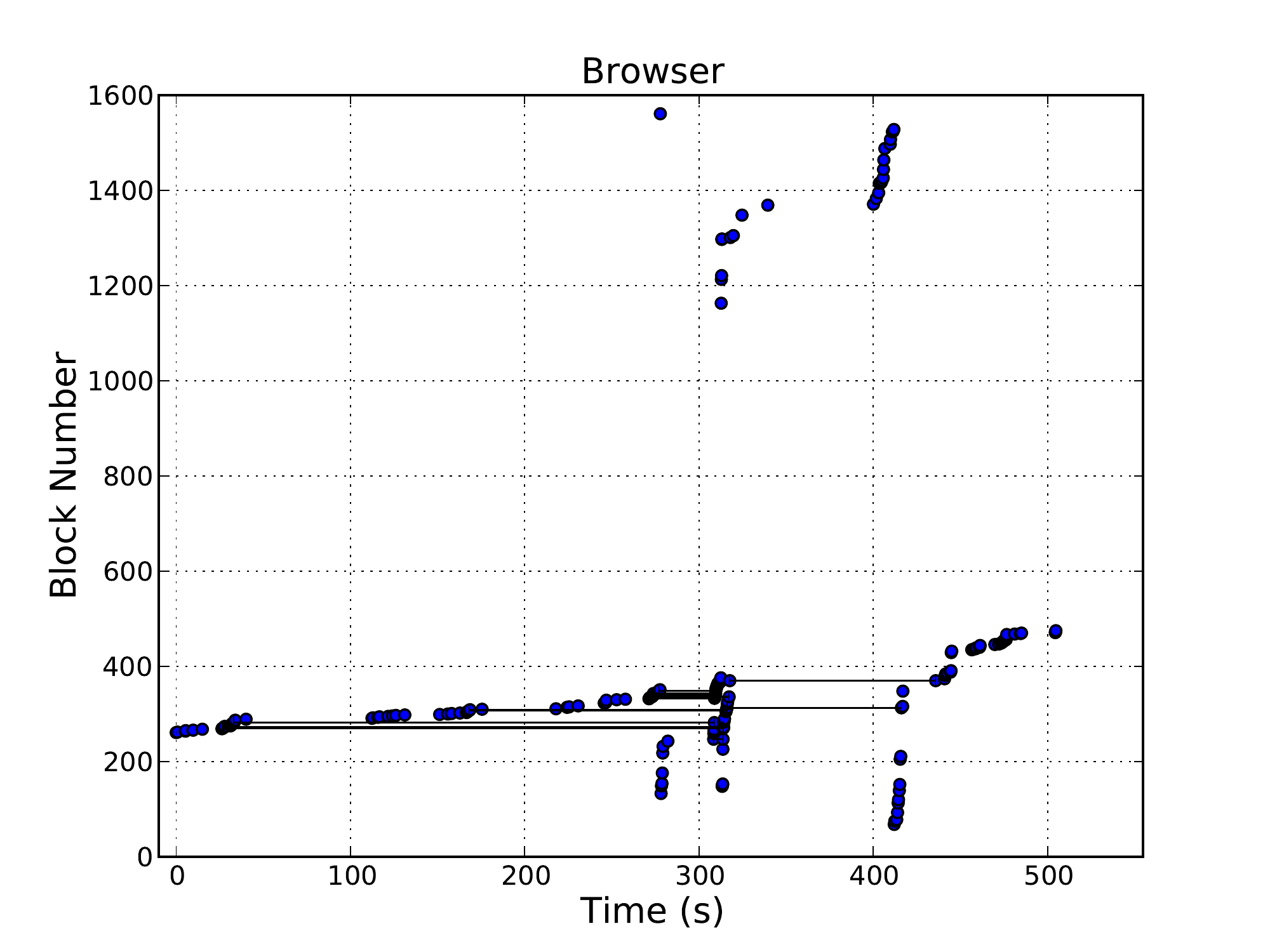}}%
  \qquad
  \subfloat[][\emph{Maps} application traces]{\includegraphics[scale=.35]{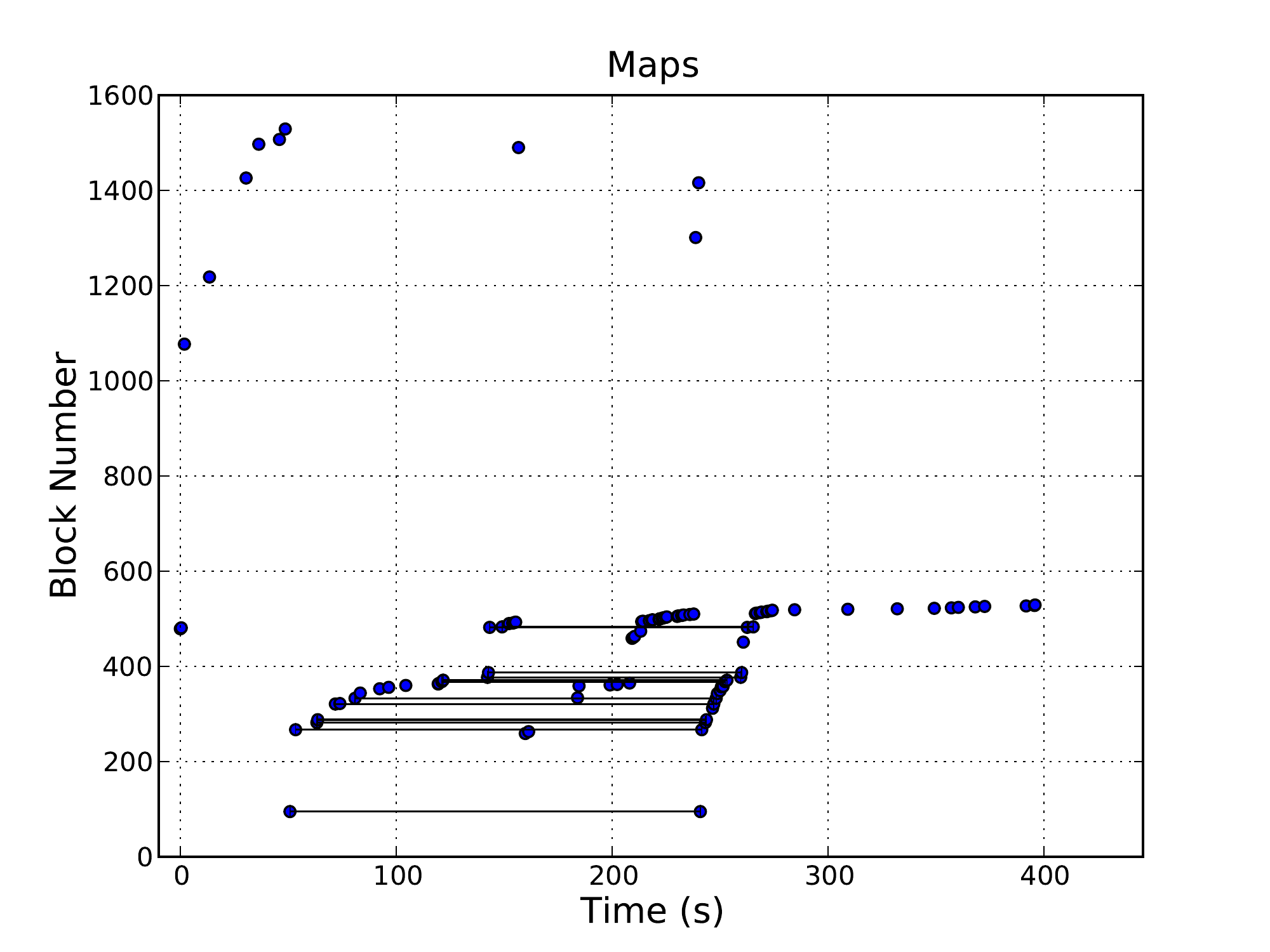}}%
  \caption{\small Snapshot of block allocations for two different
    applications. Dots represent block allocations. The only blocks
    that are reallocated are the ones between which a line is present. The
    line shows the time those blocks have been ``alive'' in
    the system before being reallocated. All other blocks are never
    reallocated throughout the test.
    \label{fig:browsermaps}\normalsize}
\end{figure}

\begin{table}[t]
\centering
\scalebox{.7}{%
\begin{tabular}{|r|c|c|c|c|c|} \hline
  & Browser & Maps & Game & Gallery & Overall \\ \hline
  Running time (s) & 504 & 395 & 300 & 240 & 1439 \\ \hline 
  Allocated blocks & 185 & 87 & 1 & 2 & 247 \\ \hline 
  Never deleted & 168 & 75 & 1 & 2 & 199 \\ \hline
  \bf Mean reallocation period (s) & \bf 54 & \bf 75 & \bf N/A &
  \bf N/A & \bf 271 \\ \hline
\end{tabular}}
\caption{\small Average reallocation period for different commonly
  used applications. The test ran for 23 minutes and allocated 304 blocks.
  \label{tab:appstats}\normalsize}
\end{table}

Statistics for each test are summarized in Table~\ref{tab:appstats}
and example traces are plotted in Figure~\ref{fig:browsermaps}. The absence of
lines after allocation of some blocks indicates that their content is still
present on the system after their deletion time.
This short usage scenario, which was executed for 23 minutes, gives an
idea of how commonly used applications write to disk.

\paragraph{Deletion Latency}
To get a better idea of deletion latency,
we used the instrumented phone daily. The experiment
lasted 670 hours, roughly 27.9 days. In total, throughout the
experiment, we recorded 20345 block allocations initiated by 73 different
\emph{writers}. A \emph{writer} could be any application including the
Android OS itself or one of its services (e.g., GPS, DHCP, compass,
etc.). The experiment's logs show that blocks
are reclaimed, on average, every 44.7 hours (the median being 44.5
hours). The worst case for block reallocation time for the experiment
is 327.7 hours. This is not surprising given the YAFFS
implementation, but it highlights the critical need for secure deletion
solutions.

\section{User-space Secure Deletion}
In this section, we introduce two solutions for secure deletion: purging and
ballooning.
Both solutions work at user-level and are designed for
the scenario where a security-conscious mobile phone
user wants to install a secure deletion 
application from an application marketplace, 
but is unwilling to install a new phone operating system.

A user-level application has limited access to the flash device. The
application cannot force the file system to perform block erasures, prioritize garbage collection of particular areas in memory, or
even know where on the device the user's data is stored. The interface to the
file system for such applications consists of the 
creation, modification, and deletion of the user's own local
files. In the next section we show a simpler solution that
requires kernel-level modifications to the Android mobile phone; here 
we propose a solution for  this highly-constrained environment.

\newpage
\subsection{Purging}

To guarantee the secure deletion of all sensitive data on a
YAFFS file system from user-space requires that we delete all the
sensitive data and then completely
fill the drive with new data. 
The fact that it must be completely filled follows from the  implementation
of YAFFS's block allocation strategy. In the worst case, we must assume a deleted chunk is 
the only deleted chunk on an erase block of otherwise live data. The block allocation
strategy first uses empty blocks, then compacts non-empty blocks by selecting the one with the fewest number of 
live chunks. In the worst case, when all other erase blocks have at least two
empty chunks, then only by
filling the drive to complete capacity are we assured that our deleted chunk is securely erased.

Purging is the operation that completely fills the file system's empty space
with a junk file, thereby ensuring that all
previously deleted data is securely erased. After filling the drive, the
junk file is deleted so that file system is again usable.
It is a rapid
operation that must be explicitly executed. This can take the form of automated
triggers,  which execute periodically when the phone is idle, whenever the browser cache is cleared, when particular apps are
closed, or upon receipt of SMS messages with self-destruction requests. It
is particularly useful for employees who are contractually obligated to delete
customer data before crossing a border.

Completely filling the drive is possible
provided the user is not subjected to disk-quota limitations, but it typically requires
garbage collecting (i.e., erasing) nearly
every erase block on the storage medium. This is because deleted chunks can occur
in any erase block that sees active use, resulting in small data gaps
throughout the file system---even chunk-aligned appends will
still erase the previous file header.

To test our hypothesis, we performed the following experiment.
We first configured an Android phone to run with our instrumented YAFFS implementation. 
We took a pristine snapshot of the phone's internal NAND memory by logging
into the phone as root,
unmounting the flash drive, and copying the raw data using \texttt{cat} from
\texttt{/dev/mtd/mtd5} (the device that corresponds to the phone's data partition) 
to the phone's external memory (SD card). The resulting file was then
copied to our PC and examined using \texttt{grep} and \texttt{hexdump}. We
wrote an arbitrary secret pattern not yet written on the device, and
obtained a memory snapshot to confirm it had been written.
We then deleted the pattern, obtained a new snapshot of the memory, and confirmed that the pattern
still remained in memory. Finally, we filled the drive to capacity with a junk
file, deleted it, and obtained
another snapshot to confirm that the pattern was now irrecoverable. The time it
took to execute the purge operation was between thirty seconds to a minute.
%During execution the system displayed a warning message that it was nearing
%drive capacity, but it disappeared after completion.

\begin{figure}[t]
\centering
\includegraphics[scale=.35]{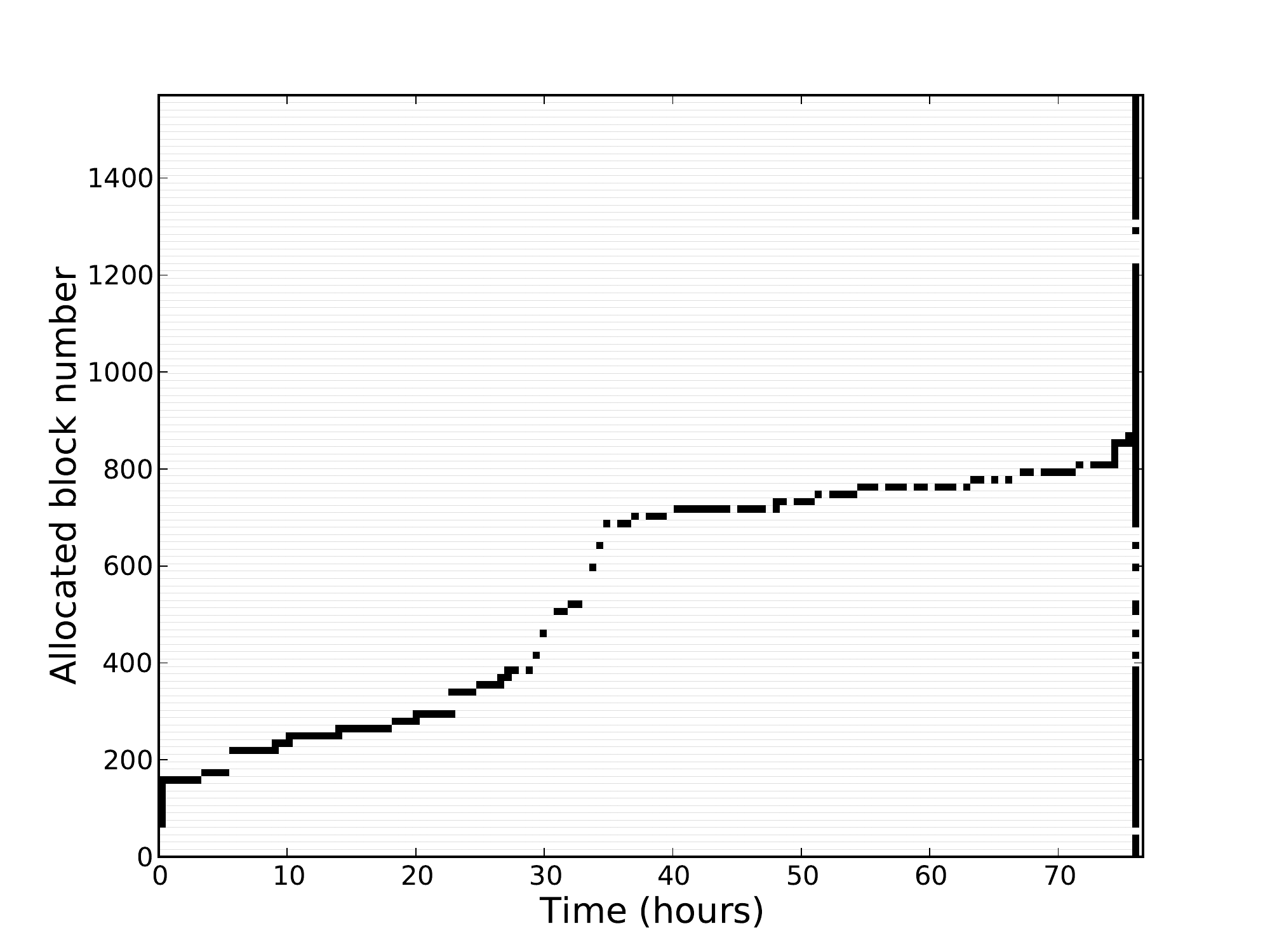}
\caption{\small Block allocation times on an Android device. The time between
two points on the same horizontal line is the block
 reallocation period. The data is a sample of live data taken from a Nexus
 One phone. At around 75
hours into the experiment, the drive was completely filled to
guarantee all previously deleted data was removed.
\label{f:alloc_gid_live}\normalsize}
\end{figure}

Figure~\ref{f:alloc_gid_live} 
shows the resulting block allocations reported by the instrumented version
of YAFFS around the time of this experiment. The X-axis corresponds to
time in hours, and the Y-axis shows the numbered erase blocks.
A small square in the graph indicates when each erase block was allocated.
At the right side, we see
the near immediate allocation of every block on the medium. This is
the consequence of filling the drive to capacity; 
YAFFS must effectively garbage collect every block 
so as to reclaim every available chunk.

\label{s:static_deletion}

\subsection{Ballooning}
In contrast to purging, which guarantees rapid
secure deletion of data from user-space, we now present ballooning, which
achieves probabilistic continuous secure deletion. Ballooning reduces
the expected time any deleted data---regardless of when it is deleted---remains accessible on a mobile phone. We begin
by looking at the time between subsequent allocations of the same flash  erase
block, which is the time  that data written on that block is accessible.

The block reallocation period in a log-structured file system is the expected time
that will elapse between allocations of a block in the file system (cf.
Figure~\ref{f:deletion_time}).
This is based mainly on two factors: the write frequency on the medium, and  the expected number of other blocks
that will be allocated before the particular block is reallocated. Were the
storage medium's size to increase tenfold, one would expect to observe a
similar increase in the block reallocation period. 

The type
of contents on the block also has an effect: 
long-term operating system files tend not to be deleted, and therefore blocks
containing only such files will not be reallocated as their
contents tend not to be deleted. Such blocks are clearly not a concern for secure
deletion, and so their existence only decreases the expected number of blocks
that will be used for non-permanent data storage.
The block reallocation period is proportional to the
expected number of blocks used for active storage and inversely proportional to the number of blocks that are
allocated per unit time.

\begin{figure}[t]
\centering
\includegraphics[scale=.35]{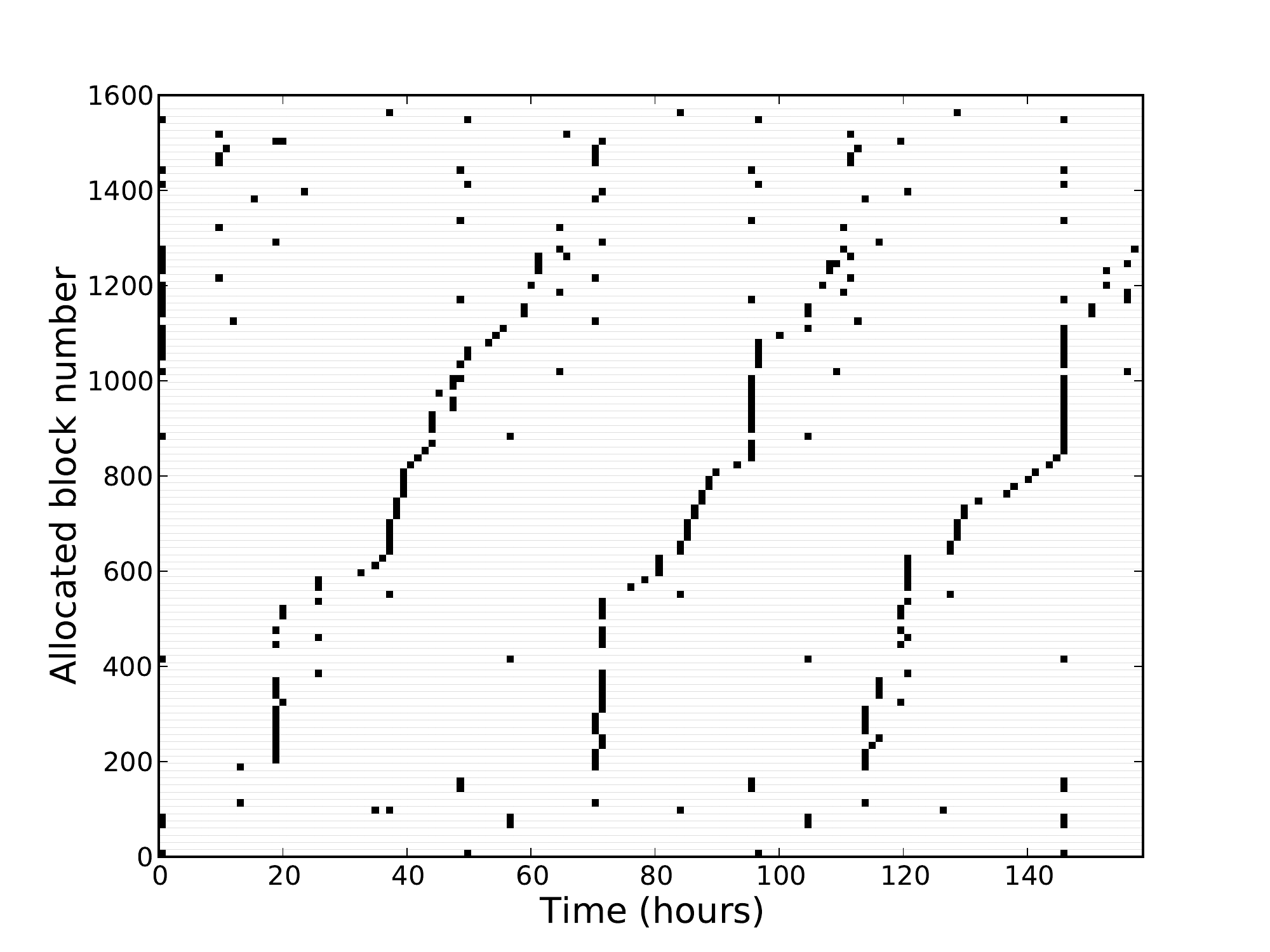}
\caption{\small YAFFS' block allocations over time  on an Android
phone. 
\label{f:block_alloc}\normalsize}
\end{figure}

The cyclic behaviour of block allocations in YAFFS is evident in
Figure~\ref{f:block_alloc} (cf. Figure~\ref{f:alloc_gid_live}), which shows
the sequence of block allocations from our collected Android
mobile phone data. 
While some noise exists, we
see that block allocation numbers generally increase over time and  wrap
cyclically, and so the block reallocation period  
is dependent on both the number of blocks and
the system-wide time between block allocations.

Our proposal is to fill the file system with junk content, thereby reducing the
block reallocation period. This reduction results from  fewer blocks being
available for allocation, and will 
thus reduce the deletion latency. As a result,
YAFFS will be forced to employ more frequent garbage collection, as the file system
will perpetually believe it is in a state of reduced capacity. 
Our application will delete the junk files
when the drive requires more space, and will regenerate them whenever
there is ``too much'' free space.

\subsection{Ballooning Application} 

\begin{figure}[t]
  \centering
  \includegraphics[scale=.4]{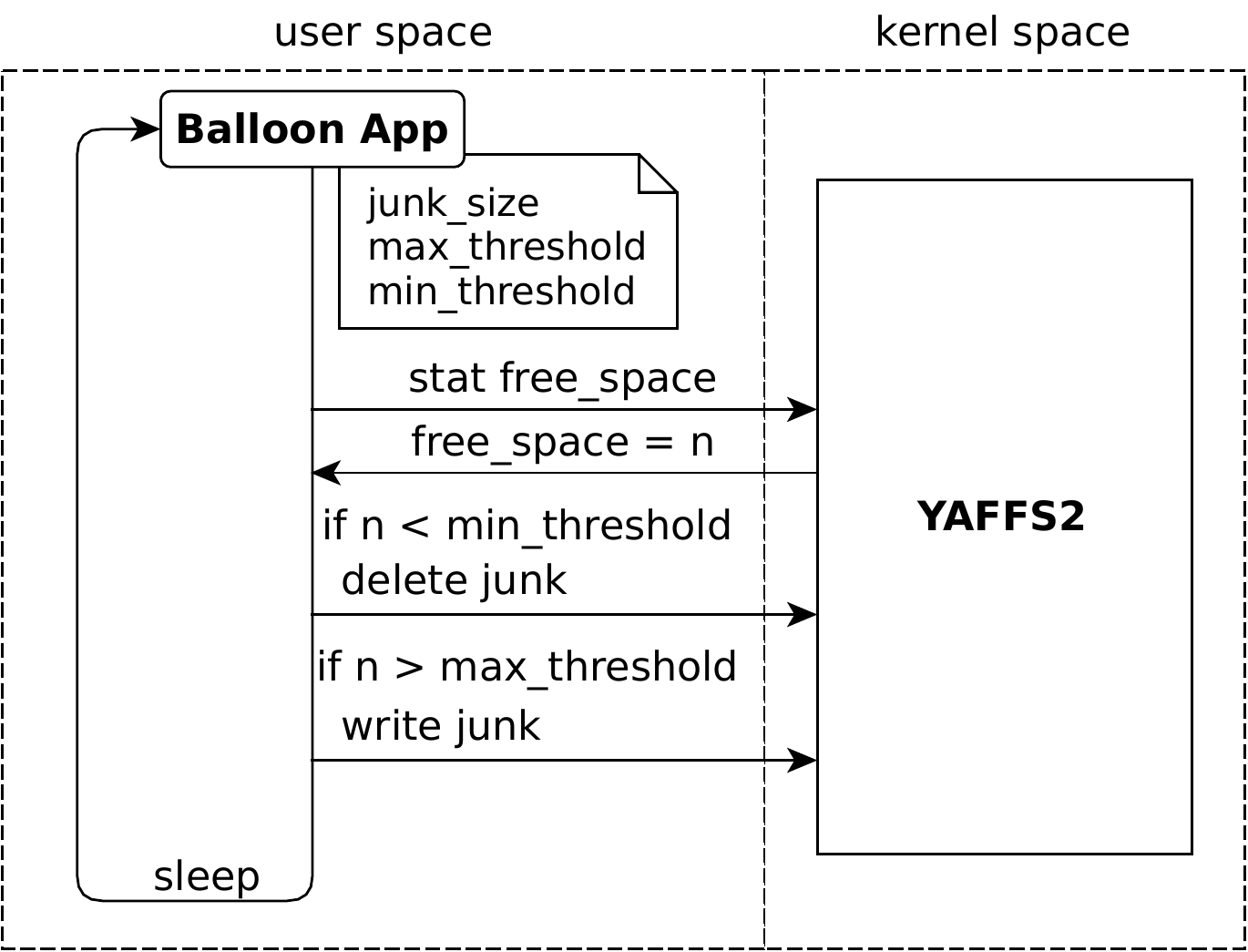}
  \caption{\small Architecture of the ballooning
  application.\label{f:application}\normalsize}

\end{figure}

The operation of our ballooning application is illustrated in Figure~\ref{f:application}. It runs
periodically on the Android phone, examining the file system (using
\texttt{stat}) to determine the number of free chunks. It creates junk files if
the free capacity exceeds the upper threshold, and deletes junk files (if
possible) when the free space drops below the minimum.
The junk files' exact size is also
parameterizable, and defined in multiples of erase blocks---deleting one junk
file will free at least one erase block for new data.

The oldest junk file is always deleted before more recent ones  to
load-balance the wear on the flash memory. Long-lived junk files can also
be erased, with new ones written, to ensure that their corresponding erase blocks will be used
for more active data storage. The new ones should be written before deleting
the old ones to ensure they reside on different erase blocks. This system is illustrated in
Figure~\ref{f:application}.

We implemented our application and ran it successfully on the Android
phone. The only permission it required was the ability to run while the phone
was in a locked state; the application also needs to specify that it will run as a
service, meaning execution occurs even when the application is not in the
foreground. The application can be installed without any elevated privileges
on the phone and operates entirely in user-space. Ballooning must 
maintain a minimum of 5\% of the blocks free to avoid
perpetual warnings about low free space.

\subsection{Experimental Evaluation}
Besides running our application on the Android phone, we collected more
statistics by simulating its
behaviour on a simulated flash drive mounted as a YAFFS file system. 
We implemented this drive in RAM using the kernel module
nandsim. 
Nandsim creates
a virtual flash device that can be mounted as any flash-based file system.
We wrote a discrete event
simulator that writes, overwrites, and deletes files on the phone's storage,
which is simply a mounted directory on our simulation computer.

Our real-life Android phone usage, described in
Section~\ref{s:deletion_android}, was used to generate all the probability distributions
for file creation, modification, and deletion.
 After a week of logging all
write activities on the phone, we computed the following two distributions for
each Android writing application: the time between successive creations of two new files, and
the type of file to create. A file type is defined by its lifetime, a
distribution over the period
between opening a file for write, a distribution over the number of chunks to write to a file each
time it is opened, and a distribution over the chunks of a file that indicates
where the writes will occur.

Additionally, we implemented  a secret writer that operated alongside the simulated
writers. It infrequently wrote  a one-chunk secret message,
waited until a new block was allocated, and deleted
the secret message.\footnote{Non-immediate deletion was done to avoid
having an erase
header get collocated with file data, since YAFFS considers erase headers as live data until all
  other traces of the file are removed from the medium.}  We logged the time before and
after we opened the file to write the secret message, and the time it was
deleted. By cross-referencing this with our block allocation information, we
determined to which blocks the secret was written, and the time when these
blocks were reclaimed thus erasing the secret.
\label{s:simulator}

In our YAFFS implementation, we
measured the rate of block allocations, which allowed us to compute the
additional cost of Ballooning as follows. The block allocation rate tells us
directly the rate that chunks are written to the flash device. Data can be written from two sources: the
actual data written by the simulator, and data copied by YAFFS's garbage
collection mechanism.  Since we are using fixed write distributions, the expected rate of writes from the simulator is 
identical between experiments. Therefore,
the observed disparity in block allocation rates reflects exactly the additional writes
that are required by our space filling application to achieve secure deletion.

\begin{figure}[t]
\centering
\includegraphics[scale=.35]{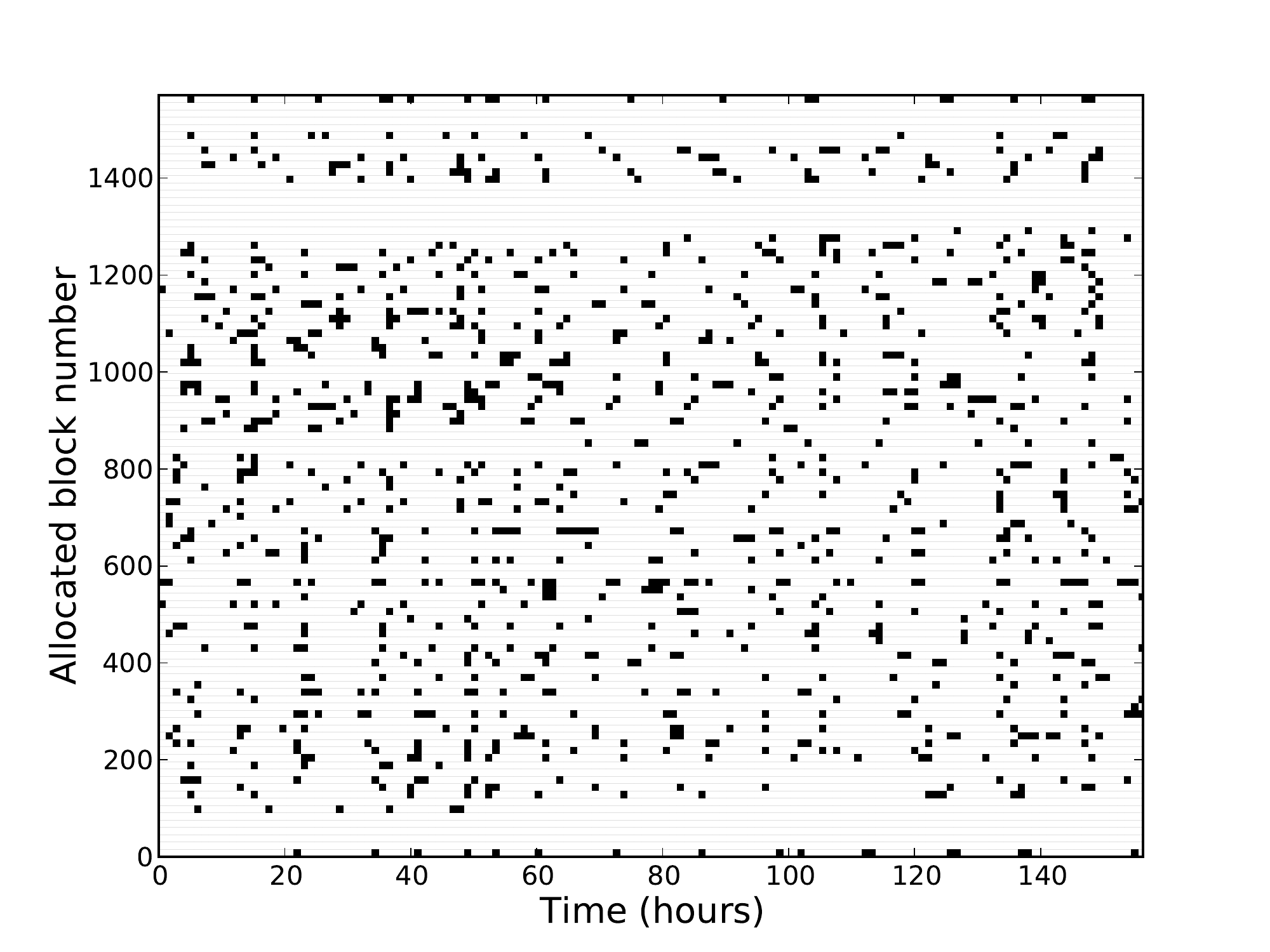}
\caption{\small YAFFS' block allocations over time on a simulated Android phone running
our ballooning application.
\label{f:bloat_alloc}\normalsize}
\end{figure}

Figure~\ref{f:bloat_alloc} (cf.~Figure~\ref{f:alloc_gid_live}) shows YAFFS block allocations when using
our ballooning application. We see that as the range of possible block
allocations shrinks considerably, the sequential allocation
strategy becomes much more erratic, and the block reallocation period
decreases. Rows in Figure~\ref{f:bloat_alloc} that contain no allocation
activity 
correspond to erase blocks that have now been assigned junk files.

To quantify the benefit of our application---that is, how promptly the secure deletion of
sensitive data occurs---we measure the expected time sensitive data remains
on the storage medium. We calculate this measurement using our secret writer
that periodically
writes one block secrets onto the medium and then deletes them. We then compute
how long the written secrets remained on the device.

\begin{figure}[t]
\centering
\includegraphics[width=0.4\textwidth]{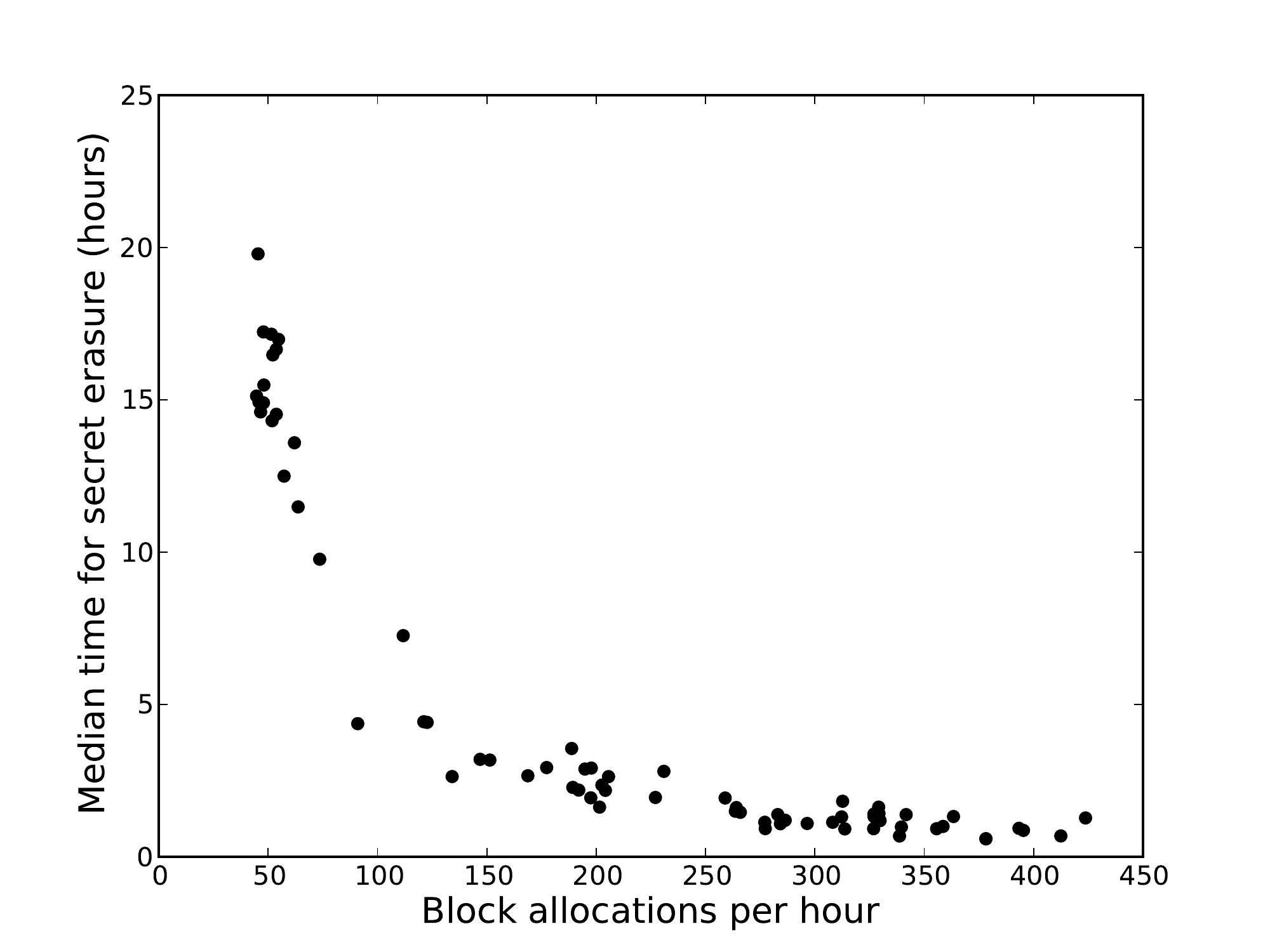}
\caption{\small Scatter plot of average block allocations per hour and median secret erasure time in
hour different ballooning simulation with a variety of parameters.\label{f:parameter_scatter}\normalsize}
\end{figure}

\begin{table*}[t]
\centering
\begin{tabular}{lrrrrr}
\hline
Free space   & \multicolumn{5}{c}{Percentile Measurements} \\
(erase  blocks)  & 1st & 50th & 90th & 95th & 100th \\ 
\hline
No ballooning  & $ 40.18 \pm 3.93 $ & $ 48.76 \pm 1.32 $ & $
55.88 \pm 2.24 $ & $ 56.99 \pm 2.37 $ & $ 58.93 \pm 2.32 $ \\

$ 250$ & $ 7.92 \pm 1.62 $ & $ 15.06 \pm 1.51 $ & $ 22.82 \pm 1.68 $ & $
23.77 \pm 1.61 $ & $ 25.03 \pm 1.29 $ \\ 

$ 50$ & $ 0.08 \pm 0.03 $ & $ 4.51 \pm 0.57 $ & $ 8.37 \pm 0.69 $ & $
 9.28 \pm 1.03 $ & $ 10.54 \pm 1.41 $ \\

$ 25$ & $ 0.06 \pm 0.05 $ & $ 2.36 \pm 0.32 $ & $ 5.24 \pm 0.81 $ & $
6.25 \pm 1.20 $ & $ 9.59 \pm 1.79 $ \\ 

$ 10$ & $ 0.02 \pm 0.01 $ & $ 1.26 \pm 0.18 $ & $ 3.14 \pm 0.27 $ & $
3.81 \pm 0.43 $ & $ 17.42 \pm 11.29 $ \\ 
Purging & $0$ & $0$ & $0$ & $0$ & $0$ \\

\hline

\end{tabular}
\caption{\small Deletion time in hours for different configuration parameters.
\label{t:erasures}\normalsize}
\end{table*}

Our application's parameters are the size of the junk files, the lower
threshold on the file system's free space when junk files are deleted, and the
upper threshold when junk files are created. 
These variables affect the total
expected free space on the partition during execution, which will be in the
range defined by the thresholds. This is typically, though not always, between the lower threshold and
the size of one junk file. The amount of free space on the drive is what
affects both deletion time and the block allocation rate. To get an idea of
how these parameters are affected, we ran our simulation for different
parameters and computed the median erasure time and block allocation rate.
Figure~\ref{f:parameter_scatter} shows the result of this experiment, which is
a scatter plot with the median deletion time on the Y-axis and block allocations
per hour on the X-axis; each point on the plot shows the results from one of our
simulations. We see from the figure that these two quantities
are inversely proportional. As the block allocations rate
increases---due to less free space and thus more frequent garbage
collection---the time secrets remain on the device decreases. 

We selected some representative configuration parameters and investigated them further.
Table~\ref{t:erasures} shows measurements of the deletion time distribution
(measured in hours) including the minimum, median and maximum measures. 
Each row of the table corresponds to a different amount of
free space (measured in the expected number of free erase blocks) as
affected by using a specific parameter set for our ballooning application. 
For each parameter set, the simulation was run eight times; the results were averaged and 
the 95\% confidence intervals were computed. 

We observe
that by leaving still 250 blocks free, corresponding to 15\% of the drive's
capacity,  we get much better secret
erasure times than if ballooning is not used, and in the extreme case of 10 free blocks,
half the secrets are deleted in an hour and a quarter.  

Since each run of the simulation uses identical write probability distributions,
we have
shown that limiting the drive's spare capacity must result in more frequent and less optimal
garbage collections. This is measured as the rate of block allocations, and in
particular, the ratio between the expected rate and the
observed rate represents the scale of the additional cost of our application.
Table~\ref{t:allocations} shows the results for block allocations (abbreviated
as allocs) per hour
using the same selected parameters for our program as with
Table~\ref{t:erasures}.

\begin{table}[t]
\centering
\begin{tabular}{lrrr}
\hline
Free space   & Block allocs & Ratio  \\
(erase blocks)  & per hour&  \\
\hline
No ballooning & $ 32.57 \pm 1.13 $ &  $1$  \\
$ 250$ & $ 52.54 \pm 4.43 $ & $ 1.61 $  \\

$ 50$ &  $ 137.47 \pm 26.92 $ & $4.22$  \\ 
$ 25$ & $ 196.00 \pm 19.03 $ & $ 6.02 $   \\
$ 10$ & $325.37 \pm 36.84$ & $9.99$  \\

\hline

\end{tabular}
\caption{\small Block allocations per hour for different configuration
parameters. The
ratio column is the proportion with regards to not using the ballooning
application.
\label{t:allocations}\normalsize}
\end{table}

We see from Table~\ref{t:allocations} that limiting the available space
significantly impacts the block allocation rate. The first step, at 250 blocks free, has
only a 61\% increase in block allocations and reasonably fast deletion.
However, achieving deletion in less than an hour requires much more
frequent block allocation. In the next section, we look at how increased block
allocations affect the device wear in terms of flash memory and
battery consumption.

\subsection{Wear and Tear}

The primary drawback of our approach is the additional wear that
increased erasures put on the mobile phone, both in terms of damage
to the flash memory and power consumption. If this approach
significantly reduced the phone's lifetime or battery life,
then it would be a concern for adoption. We therefore performed experiments to investigate this.

The additional wear is directly proportional to the increase in the block
allocation rate, and inversely proportional to the lifespan. We
convert the block allocation rate into an expected lifetime in years
using $1571$ erase blocks available on the
Android's data partition and the very conservative estimate of $10^4$ erasures per
block. (Recall that a typical flash erase block can
handle between $10^4$ and $10^5$ erasures~\cite{flashlife}).

\begin{table}[t]
\centering
\begin{tabular}{lrrr}
\hline
Free space   & Expected minimum \\
(erase blocks)  & lifetime (years) &  \\
\hline
No ballooning & 55.1 \\ 
$250$ & 34.1  \\
$50$ & 11.7 \\
$25$ & 9.1 \\
$10$ & 5.5 \\

\hline

\end{tabular}

\caption{\small Expected minimum device lifetime at various block allocation rates.
\label{t:life}\normalsize}
\end{table}

Table~\ref{t:life} shows the plot of the expected minimum lifespan of an
Android phone running continuously at varying block allocation
rates. We see that device wear is  not a concern even with
our conservative estimate of block lifetime. A device running without
our application can expect a lifespan of almost 6 decades---well beyond the
replacement period of mobile phones. Even running our application with
the most aggressive parameters still results in a lifetime of more than 5 years.
We observe there exists a trade off between wear on the device and secure
deletion, and so the user can select their desired device lifespan to
tune their security parameter.

To test if ballooning has acceptable power requirements, we
analyzed the
power consumption of write operations. We measured the battery level 
through the Android API, which gives its current charge as a
percentage of its capacity.  The experiment consisted of continuously
writing data to the flash memory of the phone in a background service
while monitoring the battery level in the foreground. We measured
how much data must be written to drain 10\% of the total
battery capacity.  We ran the experiment four times and averaged the
result. The resulting mean is within the range of $11.01\pm0.22$ GB
with a confidence of 95\%, corresponding
to $90483$ full erase blocks worth of data. Since this well exceeds
the total of $1570$ erase blocks on the Android's data partition, we
are assured our experiment must have erased the blocks as well as written to
them, thus measuring the cost of erasure.
Even using the most aggressive ballooning strategy, where $325.37$
blocks are allocated an hour, it will still take 11.5 days for the
ballooning application to consume ten percent of the
battery. Furthermore, by looking at the built-in battery use information, 
we learned that the testing
application was responsible for only 3\% of battery usage, while the
Android system accounted for 10\% and the display for 87\%. We
conclude that ballooning's power consumption is not a concern.

\section{Kernel-space Secure Deletion}
Our second solution for secure deletion is at the kernel layer, where we modify the
YAFFS file system. This models the
scenario where a mobile phone user is willing to install a custom kernel for
their mobile phone and has super-user access to the hardware.
Our goal is to provide a simple, easily auditable, and small change to the file system to achieve
secure deletion of all deleted data without additional user action.

The principle behind NAND flash programming is that an erasure sets all bits
to the value of binary one, and programming simply selects some bit positions to
instead have the value of binary zero. It is not possible to program a zero into
a one, as this operation requires erasing the corresponding erase block.
Programming a flash chunk multiple times between erasures is known as \emph{multiple
programming}.
The original version of YAFFS (YAFFS1) used multiple programming to set a
deleted flag~\cite{yaffs}. When a chunk was deleted, it was reprogrammed so this flag was
set to zero, obviating the need to perform reverse lookups in memory data
structures to determine which chunks should be copied during garbage collection. This technique was removed in
YAFFS2 to be more portable for flash memory that do not permit multiple programming.

\begin{figure}[t]
  \centering
  \includegraphics[scale=.55]{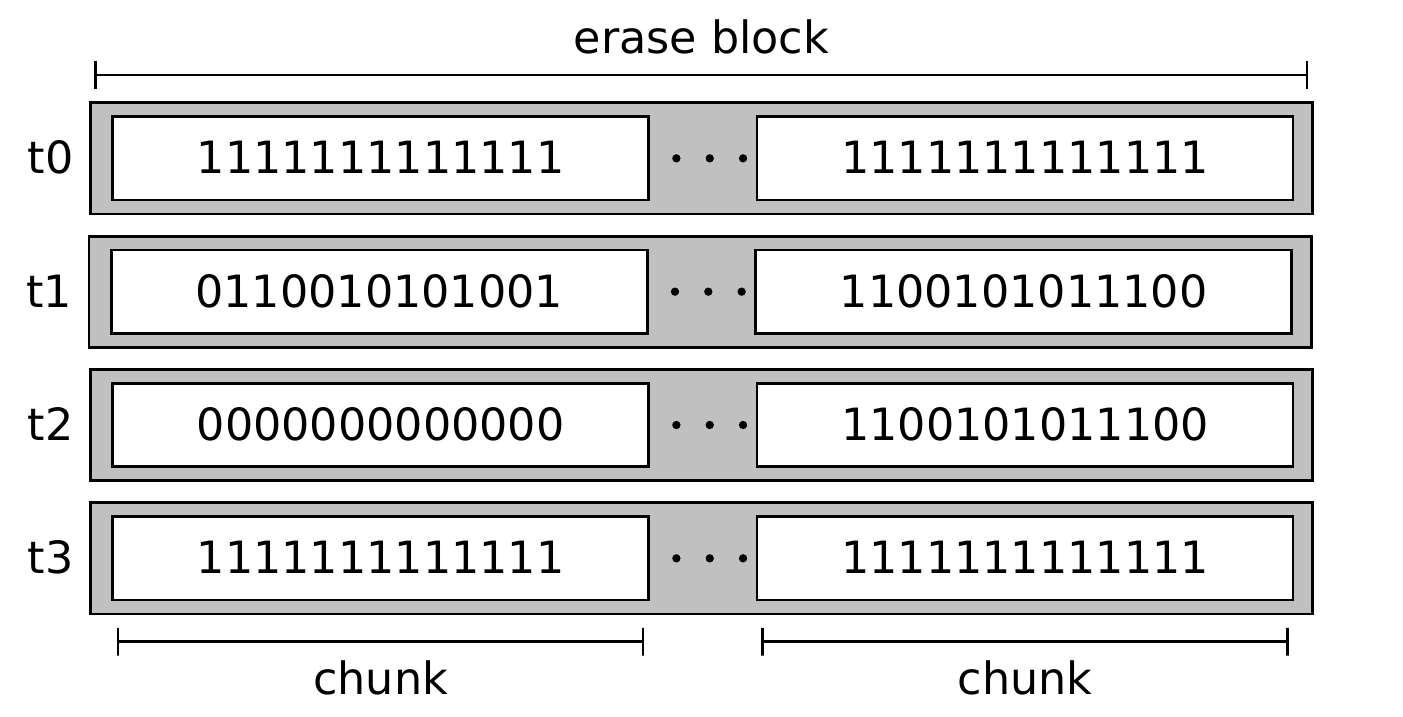}
  \caption{\small The state of an erase block at different times. $t_0$ is the
  initial state of the memory. $t_1$ is the state after some chunks have been
  written to the memory. $t_2$ shows how we can select one chunk to be written
  over with zeros, while the data remains on the other chunk. $t_3$ shows the
  memory after erasing the entire erase block, returning it to the initial
  state.\label{f:0owr}\normalsize}
\end{figure}

Our solution is to use the YAFFS1 technique of multiple programming to instead
rewrite the entire chunk's contents to zeros, thus removing the data from the
system. Since the Android's hardware supports multiple programming, portability
is not a concern for the patched kernel.
This solution requires super-user permissions to install a new YAFFS
version. It is attractive because it
requires only a tiny change to YAFFS to enable \emph{guaranteed immediate}
secure deletion without causing any additional wear on the device. 
Figure~\ref{f:0owr} shows an example of how zero overwriting removes
sensitive information.

This solution may still leak information through advanced forensic
techniques, perhaps allowing  an observer to determine how recently a gate was set
to zero, thus indicating which bits were simultaneously reprogrammed. As analog forensics on flash memory are outside the
scope of this paper, we leave this investigation as future work.

The benefit of this solution is that it requires no block erasures to delete
information. All data is immediately removed when it is no longer
needed, and the flash hardware need not erase any additional blocks
to achieve secure deletion. Therefore, there is no additional wear on the device
itself---except for the minuscule voltage required to program the
chunk to contain only zeros.

\subsection{Implementation} 
Our change to YAFFS is less than a dozen lines of C code, and is 
contained mostly in the \texttt{yaffs\_chunk\_del()} function. This function handles
the deletion of a chunk from internal memory structures, and is invoked
whenever a chunk is  no longer needed by the file system, such as deleting
a file, truncating it, or overwriting a part of it. We enhance the
method to overwrite the entire deleted chunk with zeros, using the same technique used
to set the deleted flag in the tags---which is also implemented in the same
function and used when the device is mounted as a YAFFS1 drive. The other
change is in the flash write function, where we removed a kernel oops---the
kernel equivalent of a segmentation fault---that
prohibited empty chunk tags.

\subsection{Experimental Evaluation}

We implemented  our  approach and tested its correctness by writing information,
deleting it, and then searching the raw memory for the information using
 \texttt{grep} and \texttt{hexdump}. Raw data was collected from the NAND simulator by unmounting
\texttt{/dev/mtdblock0} (the device that corresponds to our simulator) and
using the program \texttt{dd} to copy the full contents. 
Raw data was collected and examined from the Android phone using the 
technique described in Section~\ref{s:static_deletion}.

Our deletion tests consisted of creating a file with some sensitive
information and then erasing it  different ways. We tested the following: a deleted file, a file
completely overwritten, a file partially overwritten, a file
completely truncated, and a file partially truncated. The tests using partial
truncation and overwriting always erased the entire sensitive part of the file. Tests were done using
block-aligned and block-unaligned overwriting and truncation. 
We first ran our tests using the standard version of YAFFS, ensuring that the
data was still recoverable. In each test, the sensitive data is completely
erased from the file system, but remains accessible by unmounting it and
reading the raw data.  Using our modified version of YAFFS, we found that the
information was irrecoverable from both the file system and the underlying
flash medium immediately after running the deletion tests. 

A trade off with this solution is that deleting or truncating files is a linear
operation in their size, as the zero overwriting happens in the foreground. It is
also wasteful when a chunk is overwritten shortly before the entire erase
block is erased. It would be
possible to write a larger YAFFS patch that maintains a list of blocks
that need to be zero overwritten, with a sanitization daemon running in the
background; this approach is used to provide secure deletion to the ext2 file
system~\cite{ext2secdel}. Care would be needed to ensure that sanitization is
not performed if the erased block has been erased (and new data added) since
it was queued for sanitization.

\section{Related Work}

Lee et al.~\cite{treeyaffs} present a secure deletion approach for
YAFFS using encryption. They encrypt every file, and include the
corresponding encryption key in every file header written to the file system.
Secure deletion is thus achieved whenever
all the headers for a file are deleted. They propose changing the deletion
code to force deletion of header blocks containing file keys. Their approach
is elegant in that files are seamlessly encrypted and decrypted by their
proposed changes to the YAFFS file system, and it reduces the problem of rapid secure deletion to the
problem of collocating headers for the same file. They collocate headers using an
in-memory prefix-tree based on the file id, where all file headers on a leaf node
will reside on the same block. A leaf node is split into two nodes when it
is half full.

It is difficult to compare their approach with ours in experiments because
their approach was not implemented. Moreover, despite detailed algorithms, they had
not considered contracting paired leaf nodes when they are sparsely full; over
time the file system space for header data might 
sprawl as tree growth is never curtailed.
They simulated their algorithm by assuming files were modified at most twice; our examination of
Android phone data found that a third of all chunks were file headers,
suggesting much more frequent modifications. Since
their strategy is based on treating headers specially, it is important to
model them realistically. They intend to delete file header blocks
each time a file is removed; however our data indicates that Android phones delete
nearly 10000 tiny cache files a day---securely deleting each would result in
the frequent creation of bad blocks.
We suspect it would be better to delay and batch deletions, and instead of
collocating file headers based on arbitrary file id, use attributes
that may predict similar lifetimes, such as creation time or file owner.
Finally, they add secure deletion by changing an existing file
system, but do not examine how their changes effect the original design
decisions of the file system.
Device wear concerns from header collocation and
increased erasures were not discussed in their paper.

Wei et al.~\cite{wei} have considered secure deletion on flash storage in the
context of solid state drives (SDDs). An SSD makes use of a
Flash Translation Layer (FTL). This
layer allows a regular block-based file system (such as FAT) to be used on
flash memory by handling the nuances of erase blocks opaquely through the
FTL's layer of indirection. This layer has the same effect as 
a log-structured file system, where
the FTL writes new entries at empty  locations, so old entries remain
until the entire erase block can be reclaimed. They executed traditional
block-based approaches to secure deletion and determined that they do not
properly sanitize data on flash storage. They also showed alarmingly that some built-in
sanitization methods do not function correctly either. They propose to address
this concern by having 
flash hardware manufacturers make use of zero overwriting, and add it into
the FTL hardware. They state that circumventing the problem of a lack of secure
deletion requires changes in the
FTL, but depending on how the FTL is implemented, our user-level approaches may
also succeed similarly without requiring hardware changes.

\section{Discussion and Future Work}

We presented three solutions for secure deletion on YAFFS file systems:
purging and ballooning at the user-level, and zero overwriting at the kernel level.
The kernel-level solution is the most effective, and suitable for any user who
is willing to apply a small patch to their file system. Ballooning is useful
for users who wish to guard their location privacy by not having their phones
likely to record more than a user-specified time interval of location data. Purging
is useful for users who wish to be assured that some deleted data has been
securely erased from their phone.

\paragraph{Generalizing  our Approach}

Purging will work for any log-structured file system
provided both the user's disk quota is unlimited and the file system always
performs garbage collection to reclaim even a single chunk of
memory before declaring that the drive is
unwritable. 

Ballooning's utility varies with the implementation details
of the underlying file system. 
For example, JFFS is another log-structured flash file system that uses a
linear block allocation scheme. This assures ``perfect'' wear levelling; the
erasure count of any two blocks on the medium will differ by at most
one.  Consequently, filling the drive to near capacity will result in
thrashing  where all the stored data is continually being shuffled.

Zero overwriting will work for any type of file system, provided that both  the
underlying flash memory permits the second programming to occur, and the file
system will never attempt to read memory that it has already deleted. Note
that this is not the case in YAFFS1, where deleted chunks are re-read during
garbage collection to determined
if they had been marked (through reprogramming) as deleted.

\paragraph{The UBI Flash Interface}
\label{s:ubi}
Recently, Nokia has developed a new flash interface, called
Unsorted Block Images (UBI), which allows
in-place updates and removes the concerns of both wear-levelling and bad block
detection.
UBI exposes the following interface based on logical erase blocks (LEBs): read and write to a LEB, erase an
entire LEB, and atomically update the contents of an entire LEB (i.e.,
in-place edits at the erase block level).
It also allows dynamic creation of UBI partitions using unallocated
LEBs. It is neither possible for an LEB to become bad, nor is wear-levelling a
concern for LEBs.

Underlying this interface is a simple mapping from LEBs to physical erase
blocks (PEBs), where PEBs correspond to actual erase blocks on the flash
medium. Wear
monitoring is handled by maintaining a tally of the erasures at the PEB level, and
transparently remapping LEBs when necessary. Remapping also occurs  when a bad block is
detected.
Despite remapping, a LEB's numerical identifier will remain constant regardless of changes to its corresponding
PEB.

Ballooning achieved secure deletion by artificially
reducing the size of the flash partition, thus reducing the period between a
block's allocations. UBI exposes the ability to dynamically create partitions from unused
logical blocks in its block pool. It is theoretically possible for UBI to dynamically grow
or shrink the size of a partition---were it to know which blocks are not
currently being used by the file system above it, and were the file system to
know that its size is volatile.

As future work we plan to design and implement
dynamic-resizing capabilities in UBI that would also require minimal changes to
non-UBI-aware file
systems like YAFFS. These file systems view the medium as a contiguous range of
numbered blocks, along with a mapping from block numbers to states---either good
or bad. 
A UBI-enhanced YAFFS implementation might allow for dynamic resizing of its
partition size using the bad blocks map. UBI will manage how many erase blocks are given to each
partition, permitting the optimal size of
each file system to be controlled by UBI. The slack space it chooses is based
on the trade off between device wear and secure deletion. UBI would also be
able to intelligently monitor the drive, for example observing its write and erasure rate.
It may also give some partitions fewer free erase blocks than others, when
the former are being used explicitly to store sensitive files such as  encryption keys.

\newpage

\bibliographystyle{plain}
\bibliography{all}

\end{document}